\documentstyle[elsevier,12pt,epsfig]{article}

\begin{document}
\newlength{\htwid}
\setlength{\htwid}{0.48\textwidth}
\newlength{\fvskip}
\setlength{\fvskip}{-14pt}
\newlength{\cvskip}
\setlength{\cvskip}{-15pt}

\newcommand{\be}{\begin{equation}}
\newcommand{\ee}{\end{equation}}
\newcommand{\qq}{$\langle \bar q q \rangle $}
\newcommand{\rmp}[1]{{ Rev.\ Mod.\ Phys.\ }{\bf A{#1}}}
\newcommand{\npa}[1]{{ Nucl.\ Phys.\ }{\bf A{#1}}}
\newcommand{\npb}[1]{{ Nucl.\ Phys.\ }{\bf B {#1}}}
\newcommand{\prc}[1]{{ Phys.\ Rev.\ }{\bf C {#1}}}
\newcommand{\prd}[1]{{ Phys.\ Rev.\ }{\bf D{#1}}}
\newcommand{\plb}[1]{{ Phys.\ Lett.\ }{\bf{#1}B}}
\newcommand{\prl}[1]{{ Phys.\ Rev.\ Lett.\ }{\bf {#1}}}
\newcommand{\aop}[1]{{ Ann.\ Phys.\ }{\bf {#1}}}
\newcommand{\zpa}[1]{{ Z.\ Phys.\ }{\bf A {#1}}}
\newcommand{\zpc}[1]{{ Z.\ Phys.\ }{\bf C {#1}}}
\newcommand{\pr}[1]{{ Phys.\ Rep.\ }{\bf {#1}}}
\newcommand{\phr}[1]{{ Phys.\ Rep.\ }{\bf {#1}}}
\newcommand{\varro}{\rho}

\vspace*{2cm}
\title{Microscopic calculations of stopping and  flow
from 160AMeV to 160AGeV\footnote
{Supported by BMBF, DFG and GSI, ${}^b$E-mail: luke@clri6a.gsi.de, 
${}^c$Speaker at Quark Matter '96 
}}
\author{L.A.\ Winckelmann${}^{bd}$, 
S.A.\ Bass${}^{cd}$,
M.\ Bleicher${}^d$, M.\ Brandstetter${}^d$, A.\ Dumitru${}^d$, 
C.\ Ernst${}^d$, L.\ Gerland${}^d$, J.\ Konopka${}^d$, S.\ Soff${}^e$, 
C.\ Spieles${}^d$, H.\ Weber${}^d$,
C.\ Hartnack${}^f$, J.\ Aichelin${}^f$, N.\ Amelin${}^g$,
H.\ St\"ocker${}^d$ and W.\ Greiner${}^d$}
\address
{${}^d$Institut f\"ur Theoretische Physik,
         Johann Wolfgang Goethe-Universit\"at,
\\${}^e$Gesellschaft f\"ur Schwerionenforschung, Darmstadt, Germany
\\${}^f$SUBATECH, Ecole des Mines, Nantes, France
\\${}^g$Joint Institute for Nuclear Research (JINR),
Dubna, Russia}


\maketitle\abstracts{
The behavior of hadronic matter at high baryon densities is studied
within  Ultrarelativistic Quantum Molecular Dynamics (URQMD). 
Baryonic stopping is observed for Au+Au collisions from SIS up to
SPS energies. 
The 
excitation function of 
flow shows strong sensitivities to the underlying equation of state (EOS), 
allowing for  systematic studies of the EOS. 
Effects of a  density  
dependent pole of the $\rho$-meson propagator on
dilepton spectra are studied for different systems and centralities at 
CERN energies. }

\section{Introduction}

The only possibility to probe excited nuclear matter in the
laboratory are nucleus--nucleus reactions \cite{Sto86}.
In particular when two heavy ions
like Au or Pb collide most centrally, the combined system forms a zone of
high (energy) density and high agitation of the involved constituents.
The transient pressure at high density has specific
dynamic implications, such as collective sideward flow.
Hence, fundamental properties like the repulsion of the
nuclear  equation of state (EOS)
are studied
via event shape analysis of nucleons and clusters
\cite{Dos86,matt95,Part95}.
The EOS at fixed temperature 
interpreted microscopically yields
a density dependent potential modifiying the nucleon mass.
At low densities this effect is similar
as has been proposed by  
calculations on chiral limits of the Skyrme lagrangian\cite{GEB91},
the constituent quark mass\cite{KKu92} and
the chiral condensate \qq 
\cite{THa92a}.
Since \qq{}   should  relate  closely to hadron masses,
the decay of short lived vector mesons,
observed through the dilepton channel,
is suggested as a promising
experimental signal
to investigate the gradual restoration of chiral symmetry.

\section{Ultrarelativistic Quantum Molecular Dynamics}

Since many important aspects of nuclear matter
are not observable, numerical transport models are  suited
to test which assumptions are compatible to nature.
%
The present model (URQMD) \cite{uqmd,law96}
includes explicitely 50 different baryon species
(nucleon, delta, hyperon and their resonances up to  masses of 2.11GeV)
and 25 different meson species (including strange meson resonances), which
are supplemented by 
all isospin-projected states (see Table 1).
Symmetries regarding
time inversion,  iso-spin,
charge conjugation, etc.\
are implemented in a general manner,
e.g.\  all corresponding antiparticles are included and treated
on the very same (charge-conjugate) footing.
For excitations of higher masses a newly   developed 
string model is invoked.
It consistently allows for the population
of {\em all} included hadrons from a decaying string.
At low energies the dominant part of MM and MB
interactions are modelled via
$s$-channel reactions (formation and decays of resonances),
whereas  BB interactions are designed as exchange
of charge, strangeness and four momentum in the
$t$-channel.
The real part of the baryon optical potential is modeled
according to the Skyrme ansatz, including Yukawa
and Coulomb forces.


\begin{figure}[t]
\newcommand{\mysm}{\xpt}
\newcommand{\myno}{\xipt}
\begin{minipage}[c]{1.16\htwid}
\setlength{\unitlength}{3pt}
{\mysm
\begin{picture}(88,63.8)
\put(0,30.3){{\mysm
\begin{tabular}{cccccc}
 \hline  \hline  
{\myno N}&{\myno  $\Delta$}&{\myno $\Lambda$}&{\myno $\Sigma$}
&{\myno $\Xi$}&{\myno  $\Omega$} \mysm \\  \hline
 ${938}$&$ {1232}$&  ${1116}$& ${1192}$& ${1317}$& ${1672}$\\
 ${1440}$& ${1600}$& ${1405}$& ${1385}$& ${1530}$&\\
 ${1520}$& ${1620}$& ${1520}$& ${1660}$& ${1690}$&\\
 ${1535}$& ${1700}$& ${1600}$& ${1670}$& ${1820}$&\\
 ${1650}$& ${1900}$& ${1670}$& ${1790}$& ${1950}$&\\
 ${1675}$& ${1905}$& ${1690}$& ${1775}$&$$&\\
 ${1680}$& ${1910}$& ${1800}$& ${1915}$&$$&\\
 ${1700}$& ${1920}$& ${1810}$& ${1940}$&$$&\\
 ${1710}$& ${1930}$& ${1820}$& ${2030}$&&\\
 ${1720}$& ${1950}$& ${1830}$&$$&&\\
 ${1990}$&         & ${2100}$&$$&&\\
&& ${2110}$&$$&&\\
\hline\hline
\end{tabular}}}
\put(42,10.9){{\mysm
\begin{tabular}{ccccc}
\hline \hline 
$0^-$ & $1^-$ &$ 0^+$ &$ 1^+$ &$ 2^+$ \\ \hline
 $\pi$ & $ \rho$ & $ a_0$ & $ a_1$ & $ a_2$ \\
 $K  $ &$   K^*$ & $ K_0^*$ & $ K_1^*$ & $ K_2^*$\\
 $\eta$&$  \omega$& $ f_0 $&  $f_1$ & $ f_2 $\\
 $\eta'$&  $\phi $&  $\sigma$ & $ f_1'$& $ f_2'$\\
\hline\hline
\end{tabular}}}
\put(55,37){{\mysm
\begin{tabular}{c}
\hline \hline 
$ 1^-$ \\ \hline
$ \rho(1450)$ \\
$ \rho(1700)$\\
$ \omega(1420)$\\
$ \omega(1600)$\\
\end{tabular}}}
\end{picture}
}\\[9pt]
Table 1:
List of implemented baryons, mesons and their
resonances.
In addition {\em all} charge conjugate and iso-spin projected
states (and photons) are taken in and
treated on the same footing. 
\end{minipage}
\hfill
\begin{minipage}[c]{0.845\htwid}
\centerline{\psfig{figure=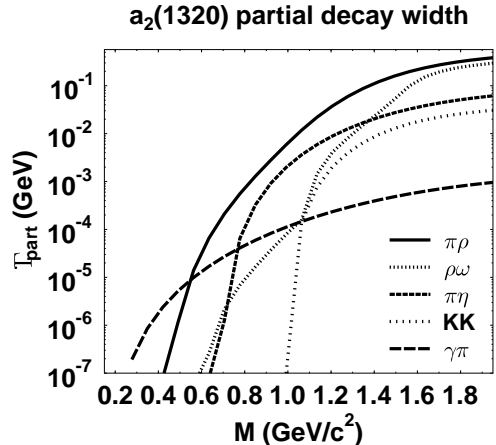,width=0.845\htwid}}
\vspace{5pt}
\vspace{\cvskip}
\caption{\label{fig:pwid}
$a_2$ partial decay rates into  specific channels.
The  average lifetime is given by the inverse of the sum.
Hence, in URQMD particles below resonance mass 
live longer,  due to shrinking phase space.}
\end{minipage}
\vspace{\fvskip}\end{figure}

\section{Creation of dense nuclear matter: stopping}

\begin{figure}[t]
\begin{minipage}[t]{\htwid}
\centerline{\epsfig{figure=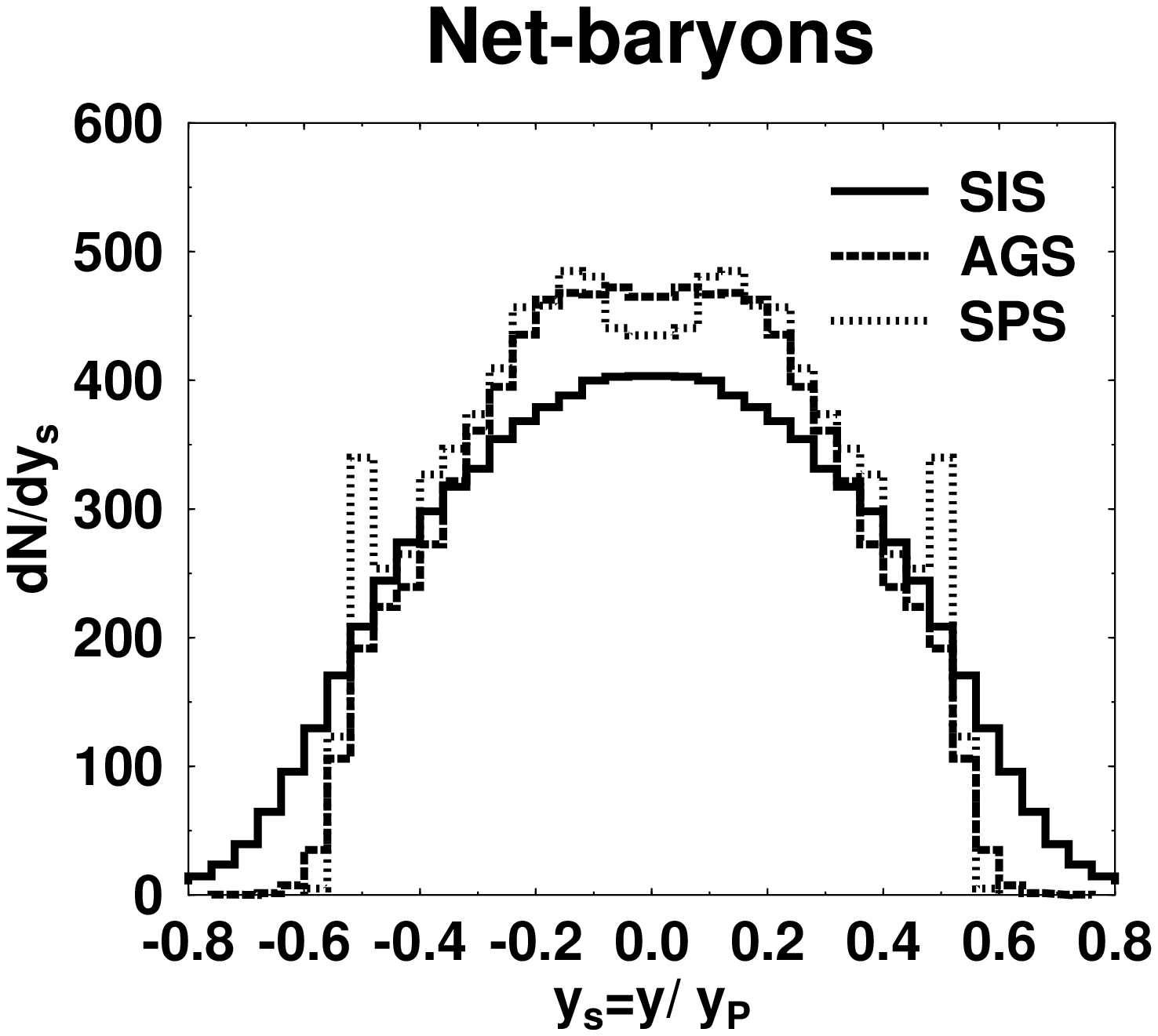,width=\htwid}} 
\vspace{\cvskip}\caption{\label{fig:stopp}Rapidity distributions for Au+Au
collisions at SIS (1$A$GeV),
AGS (10.6$A$GeV) and Pb+Pb at CERN/SPS energies (160$A$GeV).
All distributions have been scaled to the projectile rapidity in the
center of mass frame. }
\end{minipage}
\hfill
\begin{minipage}[t]{\htwid}
\centerline{\epsfig{figure=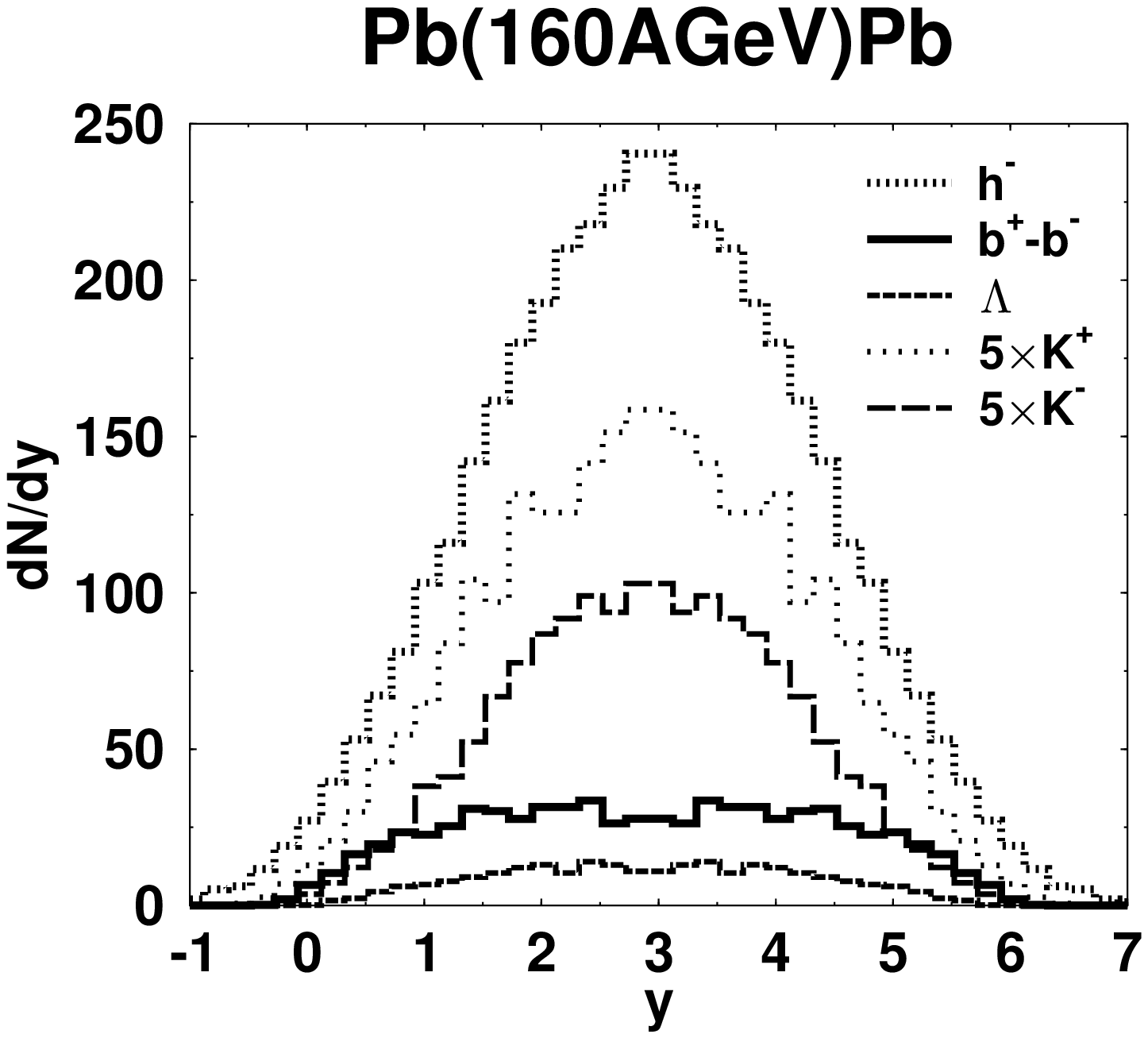,width=\htwid}}
\vspace{\cvskip}\caption{\label{fig:pbpb}
Rapidity distributions for Pb+Pb at 160$A$GeV.
The  histograms label from top to bottom: negative
hadrons  ($h^-$),   kaons ($K^-$,  $K^+$),  protons ($b^+-b^-$) and lambdas
($ \Lambda $).  The kaons are multiplied by five.
}
\end{minipage}
\vspace{1.2\fvskip}\end{figure}

Baryonic stopping is a necessary condition for the creation of hot and dense
nuclear matter. The key observable is the rapidity distribution of baryons.
It is displayed in  Figure \ref{fig:stopp} and \ref{fig:pbpb} 
for heavy systems
such as Au+Au and Pb+Pb at energies referring to
three presently used heavy ion accelerators.
In all cases gaussian rapidity distributions with peak around  
midrapidity $y_s \sim \pm 0.2$  are found.
However, the physical processes associated
show characteristic differences: The average longitudinal
momentum loss in the SIS energy regime is mainly due to the creation of
transverse momentum whereas at  AGS/SPS energies abundant
particle production
consumes a considerable amount of the incident beam energy.

At CERN/SPS energies baryon stopping is influenced
also by the formation time of strings which are
excited in  hard collisions.
In URQMD  baryons originating from a
leading constituent (di-)quark at the string edges
interact with (2/3)1/3
and mesons with 1/2 of their full cross sections during 
their formation time $\tau$.
The sensitivity on this reduction
is shown in Fig.\ref{fig:dndyform}
for the system S+S at 200$A$GeV. The default calculation (including
formation time) reproduces the data  \cite{na35} 
fairly well whereas the calculation
with zero formation time (dotted line) exhibits strongest stopping.
A calculation with zero cross section within the formation time
gives transparency.

In order to study this effect more closely the $\sqrt{s}$ distribution for 
S+S reactions at SPS energies 
is shown 
in Fig.\ref{fig:srtboth}.  
The collision spectrum
exhibits two pronounced peaks dominated by  BB collisions,
one in the beam energy range and one in the low (thermal) energy  range.
Approximately 50\% of the collisions, most of them around 
$ \sim 10 \pm 5$GeV, involve
baryons
during their formation time,
whose cross sections are reduced by factors of 2/3
(referred to as {\em di-quarks} )
or 1/3  (referred to as {\em quarks}).
This reduction leads to more transparency due to less collisions
as compared to a calculation without reduction.

\noindent
\begin{minipage}[t]{\htwid}
\centerline{\epsfig{figure=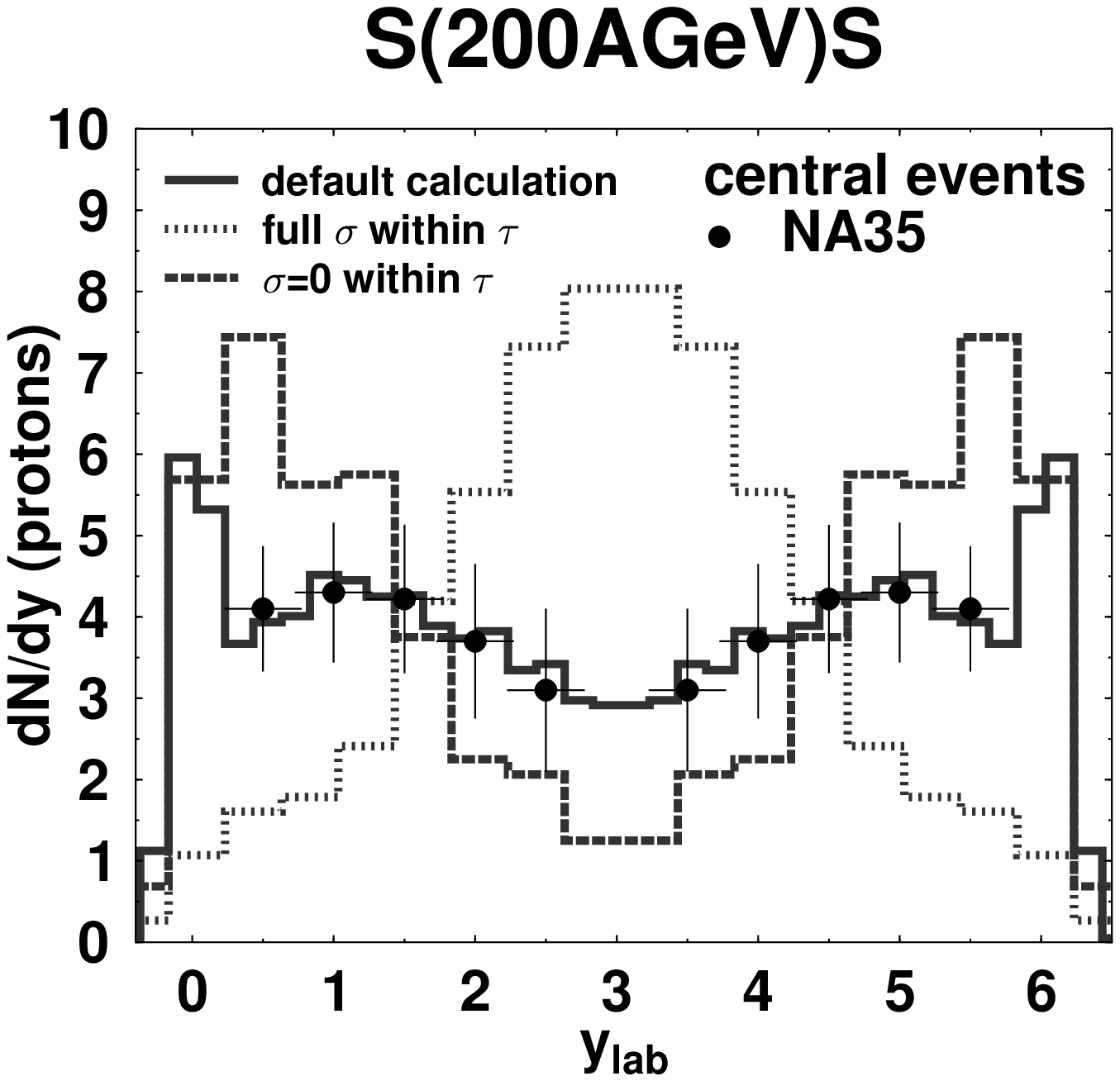,width=\htwid}}
{\figlabel{fig:dndyform}
Sensitivity of the rapidity distribution 
for S(200$A$GeV)S on the 
constituent (di-)quark cross section.} 
\end{minipage}
\hfill\begin{minipage}[t]{\htwid}
\centerline{\epsfig{figure=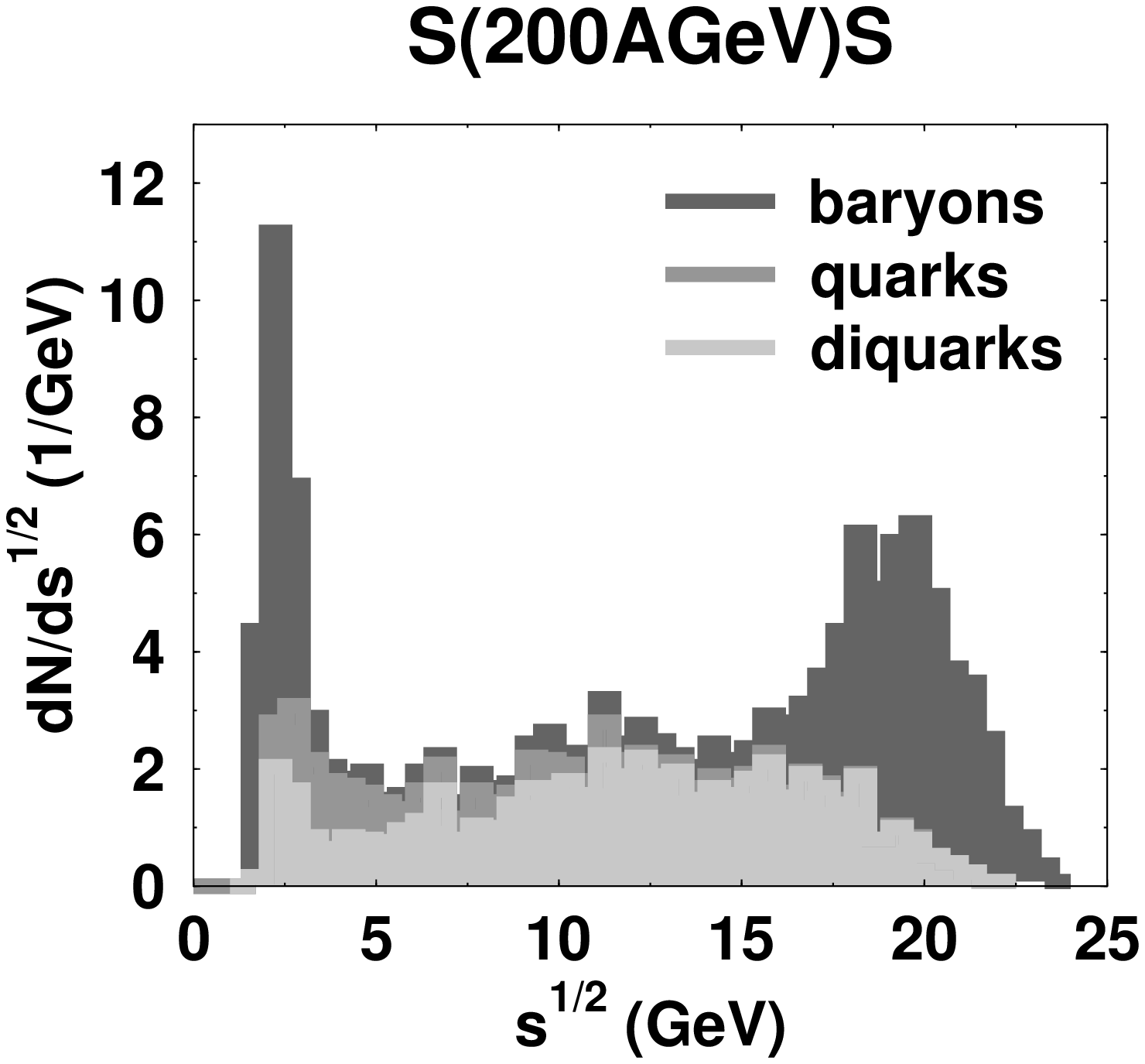,width=\htwid}}
{\figlabel{fig:srtboth}
$\sqrt{s}$ distributions for baryon baryon collisions
in central  reactions of 
S+S (right) at  
SPS energies.
}
\end{minipage}

\begin{figure}[t]
\begin{minipage}[t]{\htwid}
\centerline{\epsfig{figure=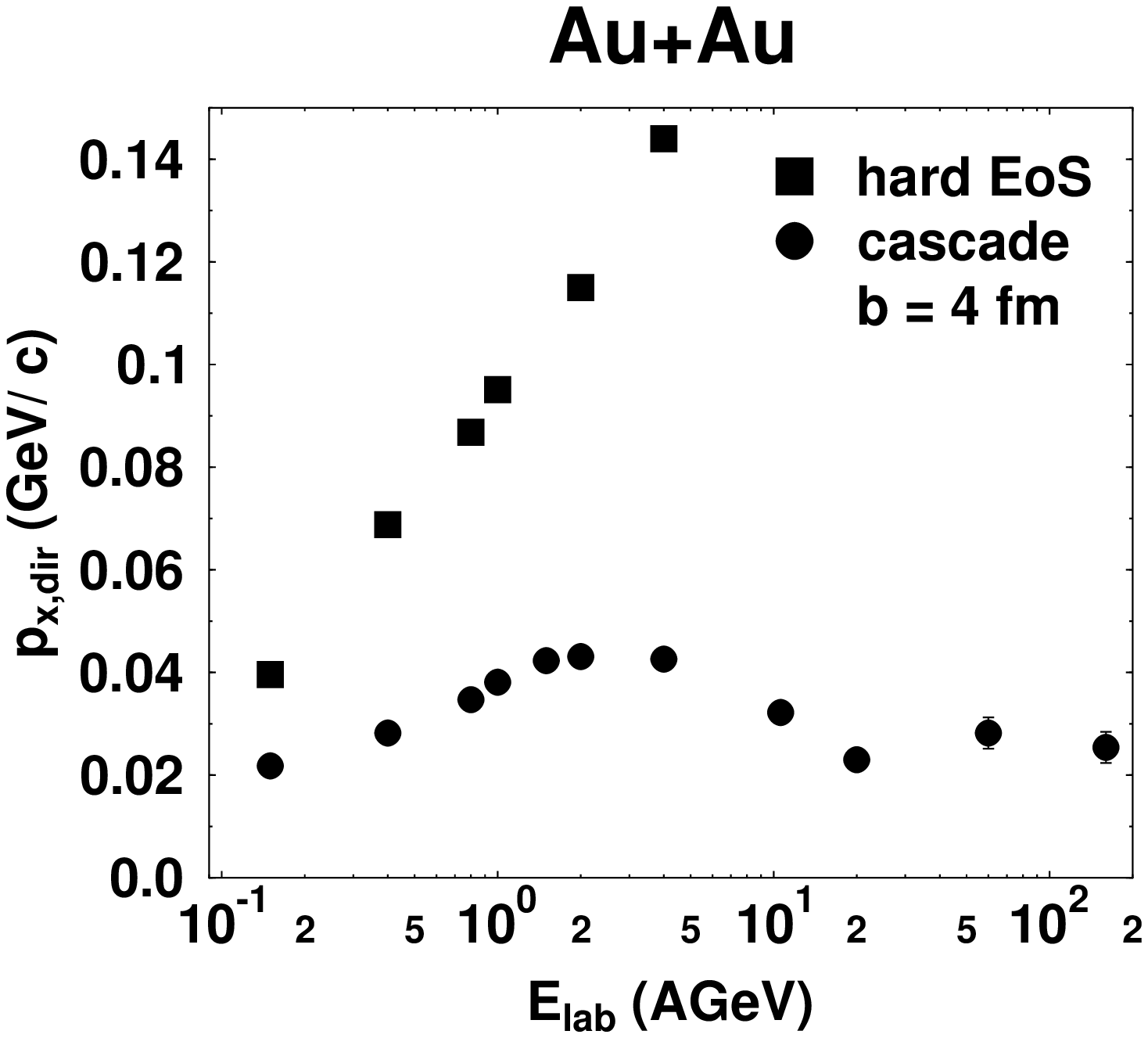,width=\htwid}}
\vspace{\cvskip}\caption{\label{fig:trans1} Excitation function of the total
directed transverse momentum transfer
px-dir for  Au+Au. URQMD calculations including a hard 
EOS (full squares) are compared to the predictions of cascade calculations
(full circles).  }
\end{minipage}
\hfill
\begin{minipage}[t]{\htwid}
\centerline{\epsfig{figure=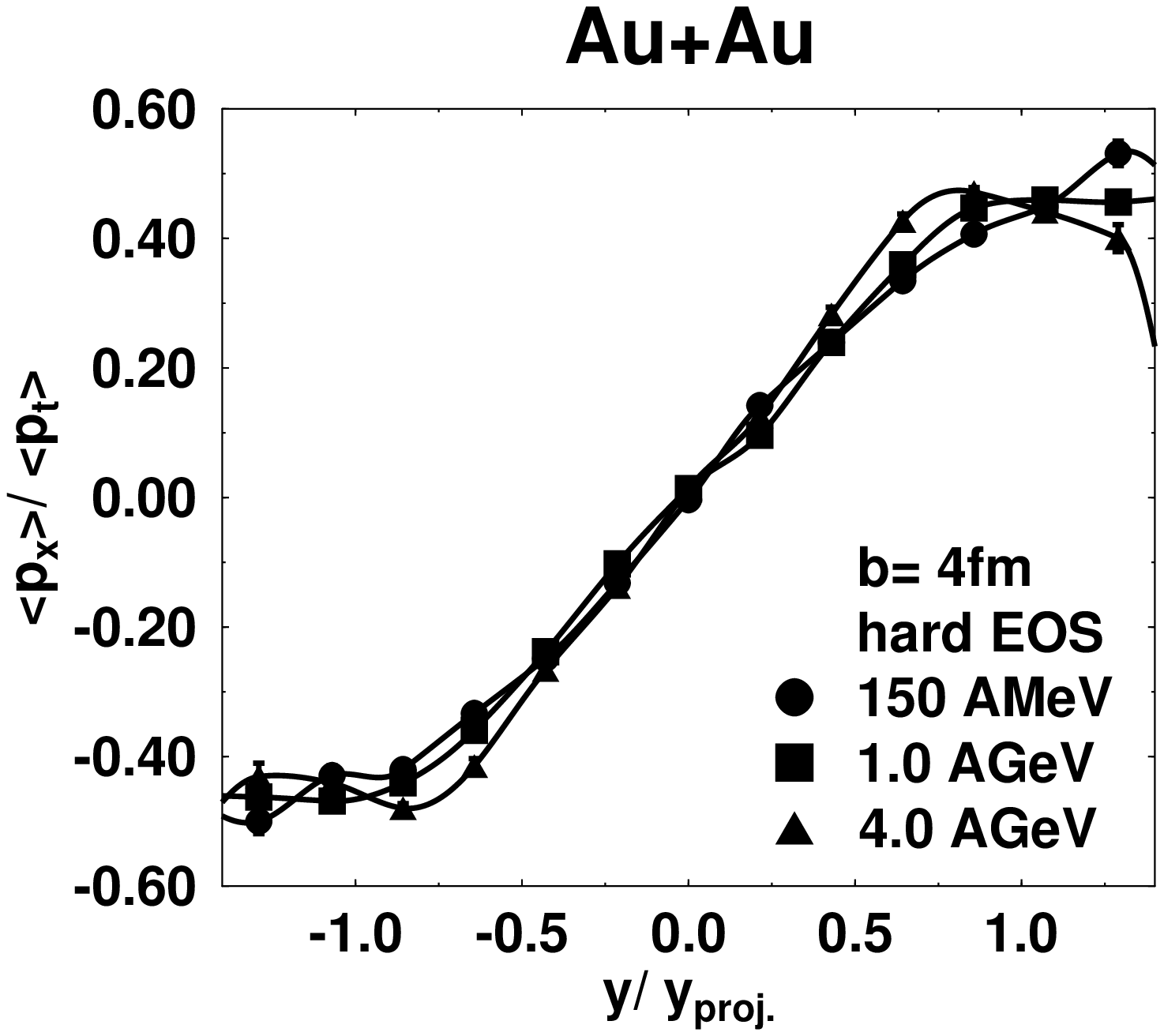,width=\htwid}}
\vspace{\cvskip}\caption{\label{fig:trans2} Mean directed transverse momentum as a
function of the scaled rapidity.
The transverse flow $<p_x>(y/y_{proj.})$  
scales with the mean transverse momentum $<p_t>$,
i.e.\  directivity does not depend on the bombarding energy. }
\end{minipage}
\vspace{\fvskip}\end{figure}

\section{Probing the repulsion of the EOS:  flow}

The creation of transverse flow is strongly correlated to the
underlying EOS \cite{Sto86}.
In particular it is believed that
secondary minima as well as the quark-hadron phase transition lead to a
weakening of the collective sideward flow. The occurrence of 
a phase transition should therefore be observable through abnormal
behavior (e.g.\ jumps) of the strength of collective motion of the
matter \cite{Rischke}.
Note that URQMD  in its present form does not include
any phase transition explicitly.
In Fig.\ref{fig:trans1} the averaged in plane transverse momentum is displayed
for Au+Au from 0.1 to 4$A$GeV incident kinetic energy.
Calculations employing a hard EOS (full squares) are compared to
cascade simulations (full circles). In the latter case only a slight energy
dependence is observed.
In contrast, the calculation with a hard EOS 
shows strong sensitivity.
Here, the integrated
directed transverse momentum  per nucleon is more than twice as high as for
the cascade calculation. This indicates the importance of a non-trivial
equation of state of hadronic matter.

The amount of directed transverse momentum scales in the very same way as
the total transverse momentum produced in the course of the reaction. 
Hence, the directivity  depends only on the reaction geometry but not on the
incident energy. This is demonstrated through Fig.\ \ref{fig:trans2},
where the mean $p_{\rm x}$ as a function of the rapidity
divided by the average transverse momentum of all particles is plotted.

\section{Temperature dependence of the EOS: photons}

\begin{figure}[t]
\begin{minipage}[t]{\htwid}
\centerline{\psfig{figure=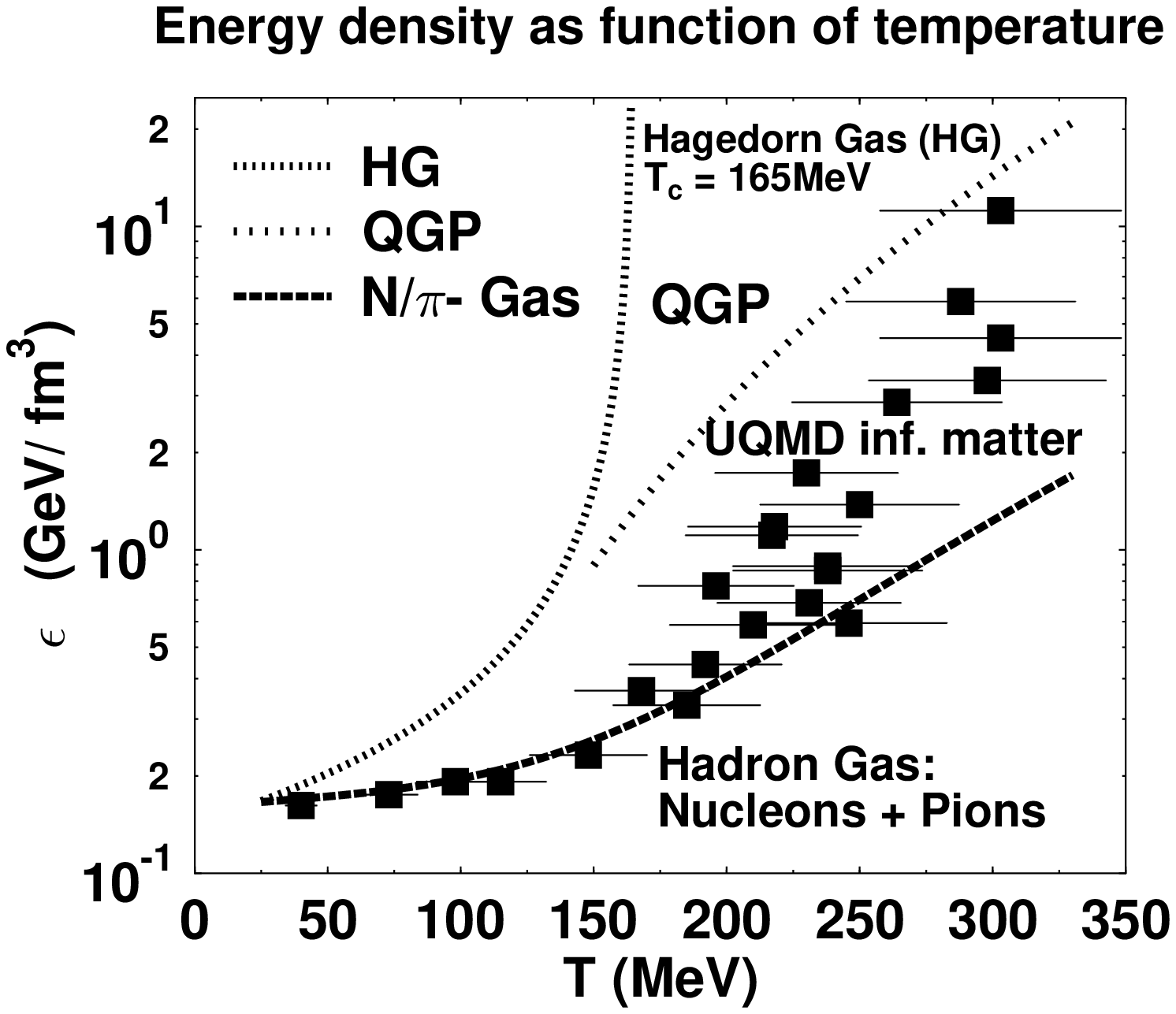,width=\htwid}}
\vspace{0.5\cvskip}\caption{\label{fig:eos}
'EOS' of infinite nuclear matter as a function of
the energy density versus temperature 
at fixed net-baryon
density of $\rho_B=0.16$/fm${}^{3}$ in URQMD (symbols).
The curves refer to analytical forms of the EOS, i.e.\
a Hagedorn-gas (top), a quark-gluon plasma (middle),
and an ideal gas of nucleons and
ultrarelativistic pions (bottom). 
}
\end{minipage}
\hfill
\begin{minipage}[t]{\htwid}
\centerline{\psfig{figure=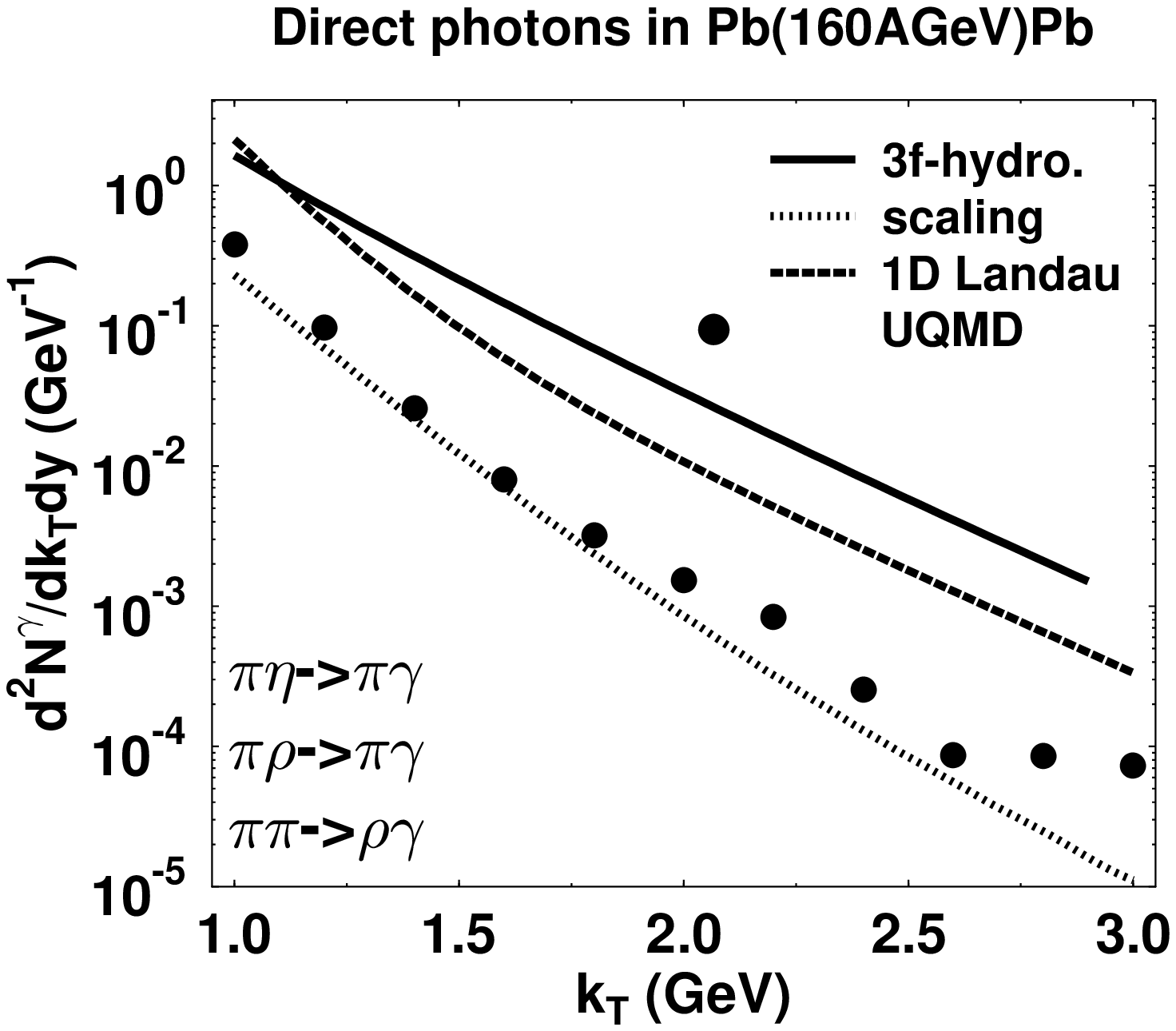,width=\htwid}}
\vspace{0.5\cvskip}\caption{Transverse momentum spectrum of directly produced photons in
Pb+Pb collisions at 160$A$GeV calculated with URQMD. 
The resulting spectrum
is compared with hydrodynamical calculations.
In all models the processes 
$\pi\eta \mapsto \pi \gamma$,  $\pi\rho \mapsto \pi \gamma$
and $\pi\pi \mapsto \rho \gamma$ are considered 
as photon sources. 
\label{photon}}
\end{minipage}
\vspace{\fvskip}\end{figure}

Semiclassical cascade models in terms of scattering hadrons
have proven to be rather
accurate in explaining experimental data.
Therefore it is of fundamental interest to extract
the equation of state from such a microscopic model, i.e.\ to
investigate the equilibrium limits and bulk properties, which are not
an explicit input to the non-equilibrium transport approach with its
complicated collision term (unlike e.g.\  in hydrodynamics\cite{Rischke,adrian}).
In Fig.\ref{fig:eos} the thermodynamic properties of
infinite nuclear matter are studied
within URQMD.

Infinite hadronic matter is simulated in URQMD by constructing
a box of 250fm$^3$ volume with periodic boundary conditions. 
According to saturation density, 
nucleons are initialized randomly in phase
space, such that a given energy density is reproduced.
After the system has equilibrated according to the
simulation with URQMD
the temperature is extracted by fitting the particles'
momentum spectra. Alternatively, the temperature can be extracted from the
relative abundances of different hadrons, e.g.\ the $\Delta/N$ ratio.

In Fig.\ref{fig:eos} the result of this procedure is compared
to various analytic forms of the EOS.
While the EOS of a Hagedorn gas and
a QGP yields energy densities $\epsilon \sim 1$GeV/fm${}^3$
at $T=150$MeV the
temperature dependence is much smaller in URQMD.
It yields about 4-5 times less energy density, being
in fair agreement with a gas composed of nonrelativistic
nucleons and ultrarelativistic pions. It remains
to be seem whether 
a reparametrization of
the resonance continuum in the Hagedorn model as suggested in
Ref.\cite{Sto81} would resolve the deviation as compared to URQMD.
On the other hand, beyond $T\sim 200$MeV
the energy density rises much faster than
$T^4$ approaching even the
QGP value of $\epsilon \sim 10$GeV/fm${}^3$
around $T=300$MeV.
This indicates an increase in the number of degrees of freedom.
It may be interpreted as a
consequence of the numerous high mass resonances and string excitations,
which seem to release constituent quark degrees of freedom (but, of
course, no free current quarks as in an ideal QGP).
Investigations
of equilibration times and relative particle and cluster abundances are in
progress. Moreover, the admittedly poor statistics have to be improved,
in order to study the high temperature behavior.

Experimentally, 
the EOS
can be accessed by measuring electromagnetic radiation
\cite{KLS}. 
In Fig.\ref{photon} the  
direct photon production from meson+meson collisions 
in Pb+Pb collisions at
160$A$GeV is shown.
Here, only 
mesons stemming from string decays are included. 
Elastic meson+meson
scattering with $\sigma_{el}=15$mb (independent of
$\sqrt{s}$) was considered. The result is compared to
calculations within the 3-fluid model \cite{adrian},
scaling and Landau expansion with $T_i=300$MeV. 

\begin{figure}[t]
\begin{minipage}[t]{\htwid}
\centerline{\psfig{figure=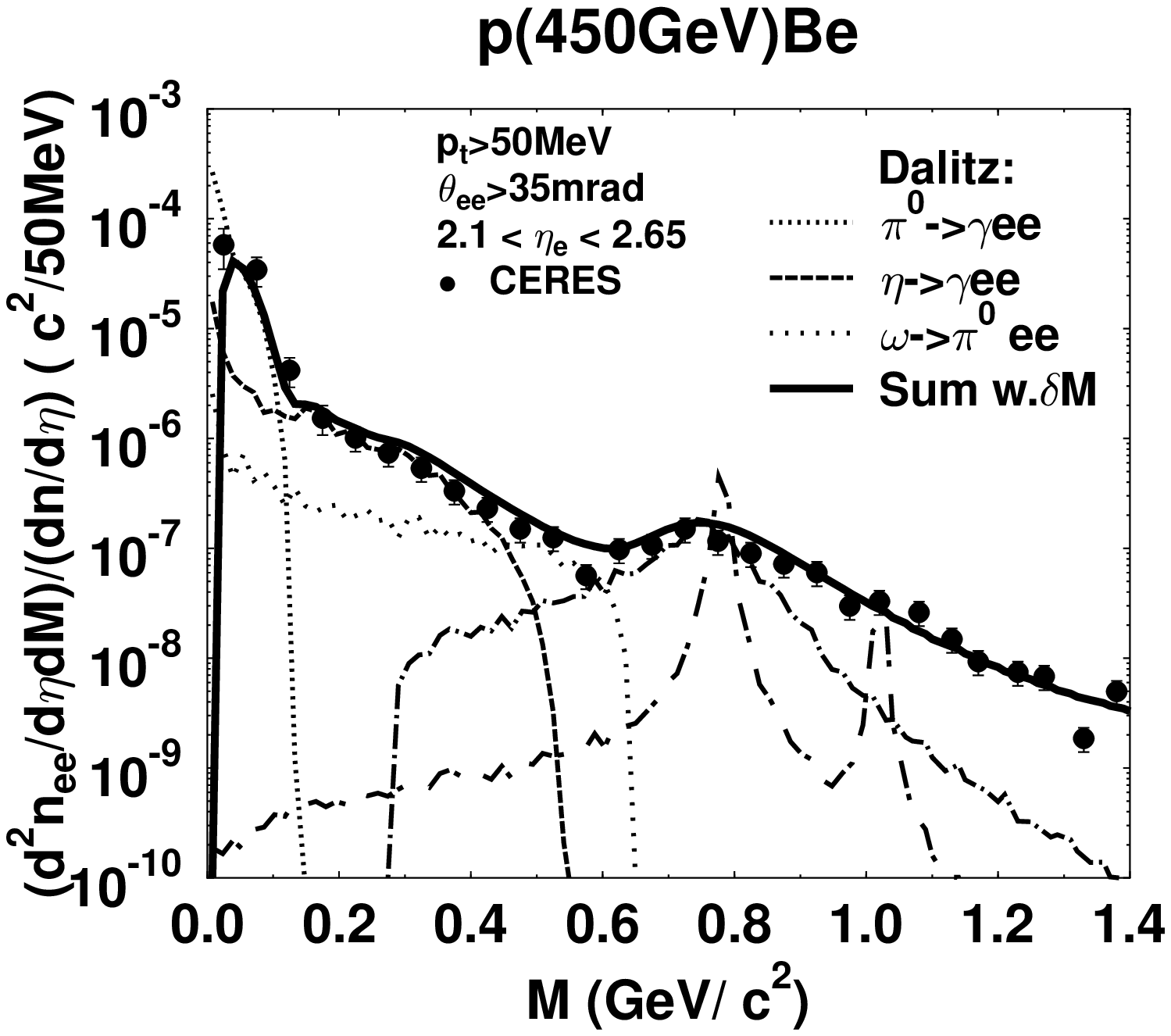,width=\htwid}}
\vspace{\cvskip}\caption{\label{pbe}
Dilepton mass spectrum for $p$+Be at 450GeV.
The calculation includes Dalitz decays and conversion
of vector mesons (see also legend for S+Au).
The sum of all contributions (solid curve) is folded with the
CERES mass resolution.
}
\end{minipage}
\hfill
\begin{minipage}[t]{\htwid}
\centerline{\psfig{figure=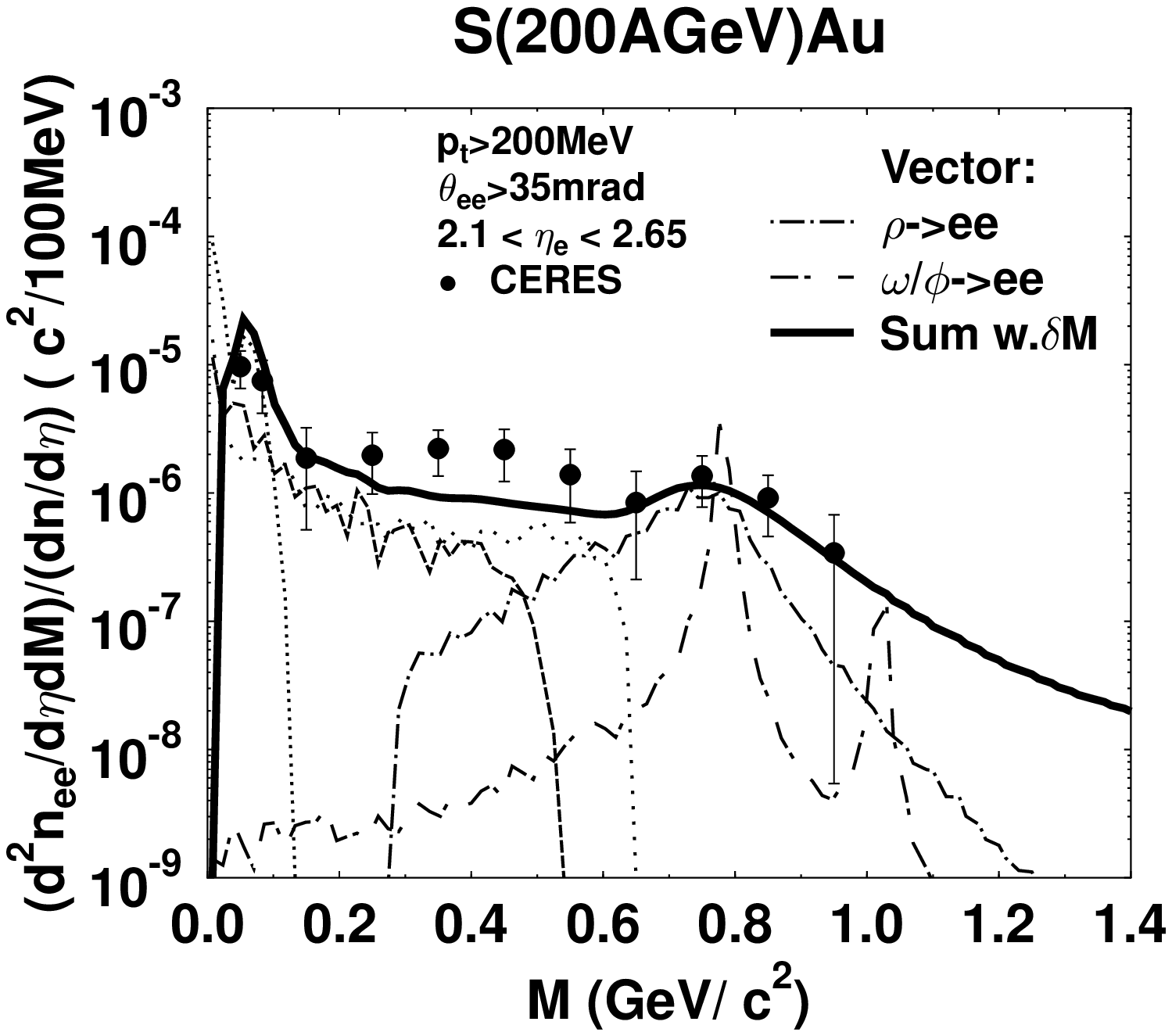,width=\htwid}}
\vspace{\cvskip}\caption{\label{sau}
Dilepton mass spectrum for S+Au at 200$A$GeV
(see also legend for $p$+Be).
Here no in-medium modifications of the $\rho$ propagator
is considered.
Around $M\sim 400$MeV two points are missed by about two standard 
deviations. 
}
\end{minipage}
\vspace{\fvskip}\end{figure}

\section{In medium masses: dileptons}

In Fig.\ref{pbe} and \ref{sau} calculations of dilepton spectra
with URQMD are show for $p$+Be
and S+Au. Dilepton sources considered here are
Dalitz decays  ($\pi^0$, $\eta$ and $\omega$)
and vector meson decays ($\rho$, $\omega$ and $\phi$).
Dalitz decays of heavier meson and baryon
resonances are included explicitely via their emission of
$\rho$ mesons  (assuming vector meson dominance).
In order to avoid double counting, the
$\rho$ mesons from $\eta$'s, and $\omega$'s are excluded
from the $\rho$ contribution.
Pion annihilation is included dynamically
into the contribution of decaying $\rho$ mesons
($\pi^+\pi^- \mapsto \rho \mapsto e^+e^-$).

\begin{figure}[t]
\begin{minipage}[c]{\htwid}
\centerline{\psfig{figure=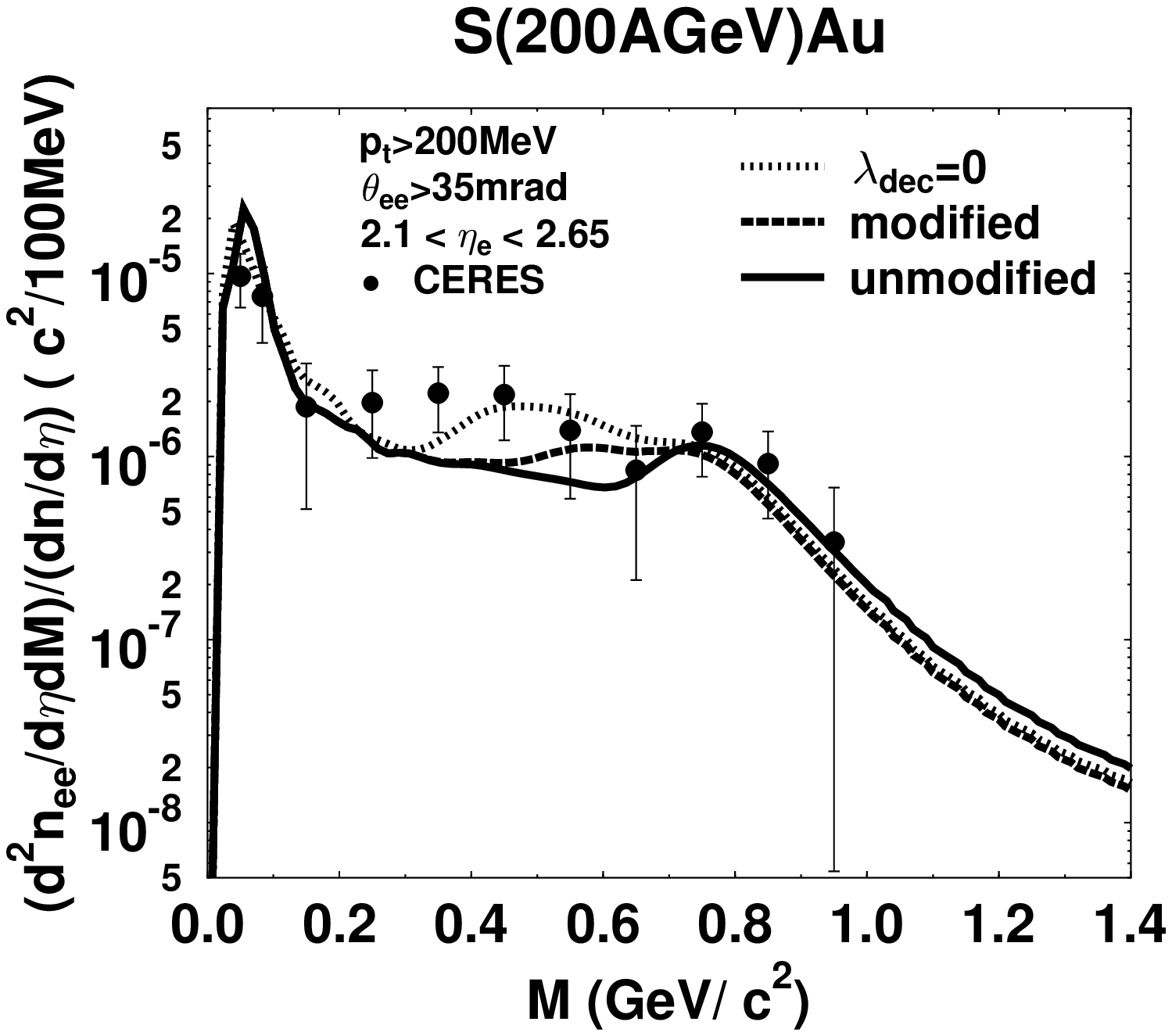,width=\htwid}}
\vspace{\cvskip}\caption{\label{sau2}
Dilepton mass spectrum for S+Au at 200$A$GeV.
The curves label simulations with pole shift according to
the creation density (top), the decay density (middle) and without
pole shift (bottom).
}
\end{minipage}
\hfill
\begin{minipage}[c]{\htwid}
\centerline{\psfig{figure=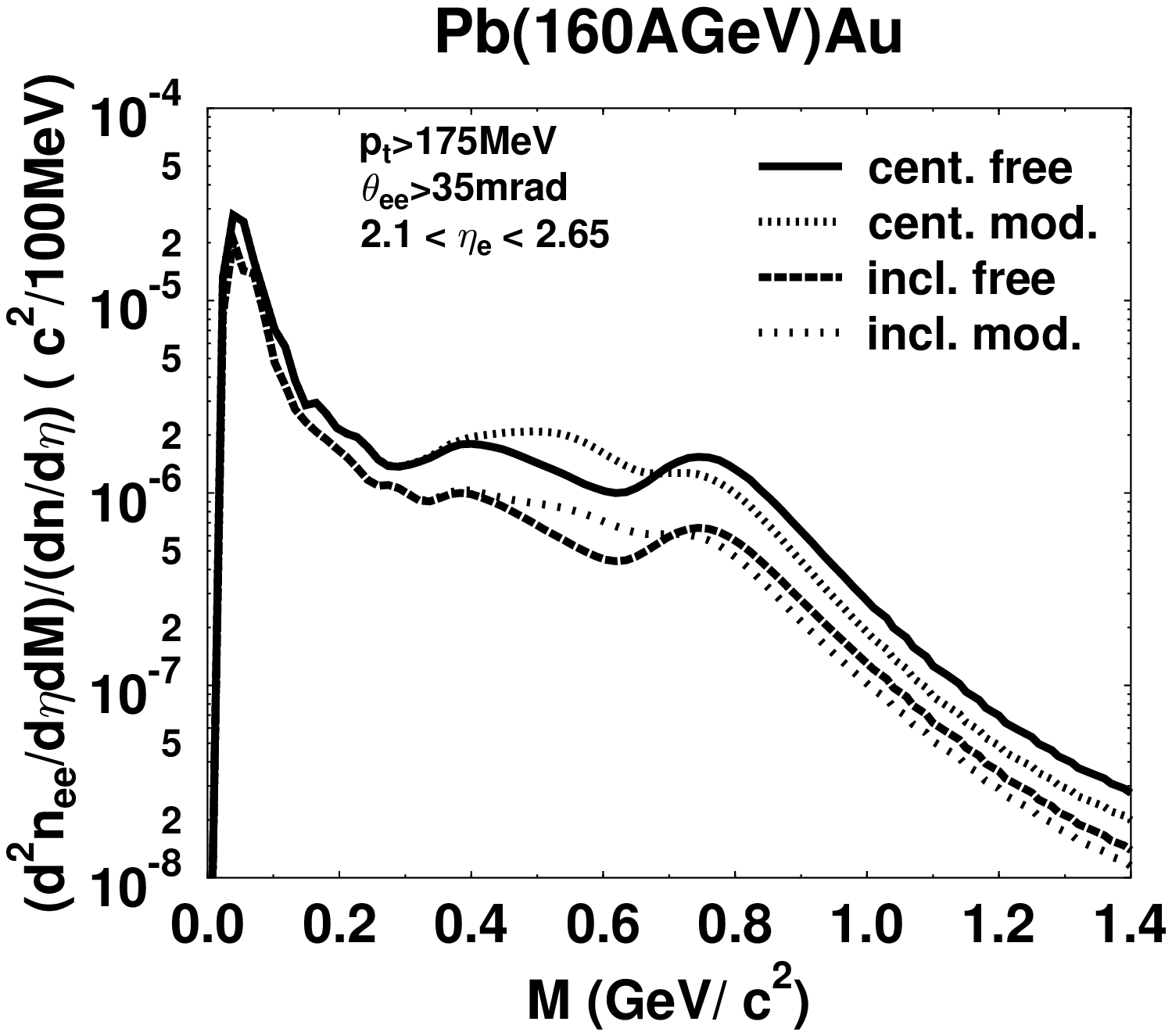,width=\htwid}}
\vspace{\cvskip}\caption{\label{pbau}
Dilepton mass spectra for Pb+Au at 200$A$GeV.
The curves label calculations 
for central (cent.) and inclusive
reactions (incl.) with (mod.) and
without a pole shift (free).
}
\end{minipage}
\vspace{\fvskip}\end{figure}

While the result for $p$+Be agrees well with the
published 
data from CERES/SPS \cite{specht},  two points around
$M\sim 400$MeV are missed by about $2\sigma$ for S+Au.
Speculations about the origin of this deviation propose  electromagnetic
bremsstrahlung,  annihilations of pions and
a modification of the $\rho$ meson propagator
due to a gradual restoration of the chiral symmetry.
The contribution of pion annihilation to the
$\rho$-peak ($\pi^+\pi^- \mapsto \rho$)
is only 40\% for S+Au.
Major additional sources
are decays of  heavy baryons $\Delta^*/N^* \mapsto N\rho$
as proposed in Ref.\cite{lw2}
and  meson resonances (see also Fig.\ref{fig:pwid}):
\begin{equation}
\left(\begin{array}{c}  \eta, ~\omega, ~\eta', ~ \phi \\
a_1, ~f_1, ~a_2, ~f_2 \\
\omega(1420), ~\rho(1450) \\ \omega(1600), ~\rho(1700)
\end{array}
\right) \mapsto \left(\begin{array}{c} \rho \gamma \\ \rho \pi \\
\rho \sigma \\ \rho \rho \end{array}
\right)~.
\end{equation}
In Ref.\cite{GEB91,THa92a} a
linear dependence of the  $\rho^0 / \omega $ pole position
as a function  of the nuclear density  $\rho $  has been  suggested:
 $m_{\rho^0}(\varrho/\varrho_0) 
	= m_{\rho^0} (0) ( 1- \lambda \varrho/\varrho_0)$.
Here  $\rho_0 $ denotes the ground state density of nuclear matter, and
 $\lambda=0.18 $, in agreement with various other calculations.
Since the restriction to low densities may not be suitable
for heavy ion collisions, the following extrapolation towards higher
densities  is  taken:
\be
    m_{\rho^0}(\varrho/\varrho_0)
     = \frac{m_{\rho^0} (0)}{ 1 + \lambda ~ \varrho/\varrho_0}
    \label{mrho}
    ~.
\ee
In Fig.\ref{sau2} an application of
Eq.(\ref{mrho}) is made to calculate
a dielectron mass spectrum for a  density dependent
vector meson pole.
In URQMD the $\rho$ meson pole position is shifted
according to the density at which the $\rho$ meson decays,
i.e.\ eventually converts
into $e^+e^-$ (middle curve). Note that this procedure
is equivalent to  a ''shining''
description  where the $\rho$ constantly emits $e^+e^-$ pairs
according to the rate $dN^{ee}/dt = \Gamma (\rho \mapsto e^+e^-)$.
This result yields only a small enhancement around $M \sim 500$MeV
as compared to the calculation without pole shift (bottom curve).
On the other hand, the data can nicely be reproduced,
if the strong (unphysical) assumption is made, that the
pole at the decay point ($\rho \mapsto e^+e^-$)
is shifted according to the creation density (upper curve).
This would be a neglection of the finite decay length.
The discrepancy of a calculation without 
decay length ($\lambda_{\rm dec}=0$fm) as compared to the result 
including the decay length is driven by two reasons:
i) The increase of the $\rho$ lifetime ($\sim 7$fm/$c$) below 
its resonance mass 
in the region $M\sim 0.3-0.5$GeV 
(where a dilepton excess in S+Au is reported\cite{specht})
lowers the decay density down
to $\langle \varrho \rangle \sim 0.2 \varrho_o$ 
for S+Au (or $0.3\varrho_o$ for Pb+Au).
ii) An enhancement of the decay length leads to an increase of 
reabsorption. Hence, the radiation path for 
$\rho \mapsto e^+e^-$ is substantially truncated.

The result for Pb+Au is shown in Fig.\ref{pbau}.
Both calculations for inclusive reactions
are in fair agreement to
the preliminary observation from CERES\cite{drees}.
For central events an enhancement due to
a nonlinear in-medium $\rho$-decay contribution
is predicted. The $\rho$ bump is located either at
$\sim 500$MeV with pole shift or at $\sim 700$MeV without
modification. Note that the deviations 
induced by the pole shifts 
are for both systems and centralities
lower than the statistical errors.
Hence, the interpretation of the experimental data 
gives the following impression: The data for light systems such as 
$p+$Be and $p+$Au (see also Ref.\cite{ko})
as well the data for heavy systems both for inclusive   and central
reactions are reproducible whithout mass shifts. In contrast the 
central data for S+Au exceed the URQMD calculation
around $M\sim 0.4$GeV by about two standard deviations

\section{Summary}

Studies of the equation of state and consequences
of gradual restoration of the chiral symmetry
are presented using a microscopic phase space model
including 75 hadron species and strings.
The directed transverse momentum 
shows strong 
sensitivities on the underlying EOS:
There is only a small increase in the cascade calculation,
whereas it scales linearly with the
average transverse momentum for a hard equation of state.
Hence, measurement of the excitation function of
the transverse directed flow allows for a systematic study
of the EOS.

The increase of the $\rho$ lifetime  at
masses $M\sim 0.4$GeV where
a dilepton excess in S+Au is reported\cite{specht}
lowers its average $e^+e^-$ emission density substantially. 
Therefore it seems that the density dependence 
of the $\rho$ pole alone 
does not suffice to explain the dilepton excess in S+Au.
Hence, other effects  -- such as the temperature dependence
of \qq{} or additional sources -- 
might be required.

\setlength{\listparindent}{2cm}

\end{document}